\documentclass[preprint,5p,sort&compress]{elsarticle}
\usepackage{epsfig,cite}
\usepackage{amsmath}
\usepackage{amssymb}
\usepackage{amsfonts}
\usepackage{latexsym}
\usepackage{wasysym}
\usepackage{url}
\usepackage{hyperref}
\usepackage{bm}
\usepackage{graphics}
\usepackage{rotating}
\usepackage{booktabs}

\journal{Physiscs Letters B}
\begin{document}

\begin{frontmatter}

\title{Small- and large-$x$ nucleon spin structure \\
from a global QCD analysis of polarized Parton Distribution Functions}
\author[unige]{E.~R.~Nocera\corref{cor1}}
\ead{emanuele.nocera@edu.unige.it}
\cortext[cor1]{Corresponding author}
\address[unige]{Dipartimento di Fisica, Universit\`a di Genova and
INFN, Sezione di Genova,\\ Via Dodecaneso, 33 I-20146 Genova, Italy}

\begin{abstract}
I investigate the behavior of spin-dependent parton distribution functions 
in the regions of small and large momentum fractions $x$. 
I present a systematic comparison between predictions for relevant observables 
obtained with various models of nucleon spin structure and a recent global 
analysis of spin-dependent distributions, {\tt NNPDFpol1.1}.
Together with its unpolarized counterpart, {\tt NNPDF2.3}, they form a 
mutually consistent set of parton distributions. Because they include
most of the available experimental information, and are determined with a
minimally biased methodology, these are especially suited for such a study.
I show how {\tt NNPDFpol1.1} can discriminate between different theoretical 
models, even though {\tt NNPDF} uncertainties remain large near the endpoints
$x\to 0$ and $x\to 1$, due to the lack of experimental information.
I discuss how our knowledge of nucleon spin structure may be improved 
at small $x$ by future measurements at an Electron-Ion Collider, and at 
large $x$ by recent measurements at Jefferson Lab, also
in view of its $12$ GeV 
upgrade.
\end{abstract}


\end{frontmatter}

\paragraph{Introduction}
{
The behavior of spin-dependent, or polarized, Parton Distribution Functions
(PDFs) at small and large momentum fractions $x$ has been recognized for
a long time to be of particular physical 
interest~\citep{Bass:2004xa,Kuhn:2008sy}.
On the one hand, the small-$x$ region is pivotal for revealing 
new aspects of the nucleon picture depicted by Quantum Chromodynamics (QCD),
related, for instance, to PDF evolution.  
On the other hand, the large-$x$ region is definitive of hadrons: indeed,
all Poincar\'{e}-invariant properties of a hadron, like flavor 
content and total spin, are determined by valence quark PDFs
in the region $x\gtrsim 0.2$, where they are expected to dominate. 
Above all, an accurate knowledge of polarized PDFs over a broad 
range of $x$ values is required to reduce the uncertainty 
with which the first moments of polarized distributions and structure 
functions can be determined. This is relevant for testing various sum 
rules~\citep{Bjorken:1966jh,Ellis:1973kp,Burkhardt:1970ti,Efremov:1996hd} 
and potential SU(3) flavor-symmetry breaking~\citep{Cabibbo:2003cu},
and finally for assessing quark and gluon contributions to
the nucleon spin. 

Several recent studies~\citep{Hirai:2008aj,deFlorian:2009vb,Leader:2010rb,
Blumlein:2010rn,Ball:2013lla,Arbabifar:2013tma,Jimenez-Delgado:2013boa,
deFlorian:2014yva,Nocera:2014gqa} have presented a determination of 
polarized PDFs, along with an estimate of their uncertainties. 
These parton sets differ in the choice of data sets, details of the QCD
analysis (such as the treatment of heavy quarks or higher-twist corrections)
and the methodology used to determine PDFs, including the form of PDF
parameterization and error propagation (for details, see {\it e.g.}
Chap.~3 in Ref.~\citep{Nocera:2014vla}).

Despite remarkable experimental efforts, the kinematic coverage of the 
available data sets to be included in global analyses is still rather limited.
Specifically, the accessed range of parton momentum fractions is 
roughly $10^{-3}\lesssim x \lesssim 0.5$: thus, a determination of
polarized PDFs outside this region would be very much prone to the
functional form used for extrapolation.  

Various models have been developed for predicting the polarized 
PDF behavior at small and large $x$. Computations based on different 
models often lead to rather different expectations for some polarized 
observables. A way to discriminate among models, and eventually test 
their validity, is to compare predictions for such observables, 
obtained within either a given model or a reference parton set.
The latter should be determined from a global QCD analysis of 
experimental data.

The goal of this paper is to present such a comparison in a systematic way,
separately for small- and large-$x$ regions.
I will also discuss how our knowledge of nucleon spin structure may be 
improved, respectively at small and large $x$, by future measurements 
at a high-energy polarized Electron-Ion Collider (EIC)~\citep{Accardi:2012qut}, 
and  by recent measurements at Jefferson Lab (JLAB),
also in view of its $12$ GeV upgrade~\citep{Dudek:2012vr}.

In order for this study to be effective, the choice of the reference PDF set 
is crucial. On the one hand, it is highly desirable that most of the available 
experimental information is included in it, so that
the unknown extrapolation region is reduced as much as possible.
On the other hand, it is fundamental that a minimal set of theoretical 
assumptions and a procedure which allows for a faithful estimate of PDF 
uncertainties are used. 

Among all PDF sets available in the literature, the {\tt NNPDF}
parton sets are those which best fulfill the aforementioned requirements
(and possibly the only). Hence, these will be used in this study: specifically,
{\tt NNPDFpol1.1}~\citep{Nocera:2014gqa} for polarized PDFs and 
{\tt NNPDF2.3}~\citep{Ball:2012cx} for the unpolarized,
whenever also these will be needed.

Concerning the experimental information included in these parton sets,
a large amount of high-precision Hadron Electron Ring Accelerator (HERA) 
and Large Hadron Collider (LHC) data are taken into account in {\tt NNPDF2.3}, 
while polarized hadron collider data sets, specifically jet and $W$-boson 
production provided by the Relativistic Heavy Ion Collider 
(RHIC), are used in {\tt NNPDFpol1.1}.
Some of these data are missing in other global unpolarized/polarized analyses 
so far. In the unpolarized case, only a subset of LHC data 
are included in recent PDF determinations~\citep{Ball:2012wy}. 
In the polarized case, $W$-boson production data are included only in 
{\tt NNPDFpol1.1}, and jet production data are included only in 
{\tt NNPDFpol1.1} and in the determination of Ref.~\citep{deFlorian:2014yva}. 
Other recent analyses are based on inclusive Deep-Inelastic Scattering
(DIS) data solely~\citep{Hirai:2008aj,Blumlein:2010rn,Ball:2013lla,
Jimenez-Delgado:2013boa},
or on inclusive and semi-inclusive DIS (SIDIS) 
data~\citep{deFlorian:2009vb,Leader:2010rb,Arbabifar:2013tma}.\footnote{Note
that pion production data from RHIC, not included in {\tt NNPDFpol1.1}, 
are also taken into account in the determinations of 
Refs.~\citep{deFlorian:2009vb,deFlorian:2014yva}.
It was argued in Ref.~\citep{Nocera:2014gqa} that these data may have a limited
impact though.}
SIDIS data sets are not included in {\tt NNPDFpol1.1}. 
However, these bring in information mostly on quark-antiquark
separation at medium-$x$ values, and they are expected to be of
limited importance in the small- and large-$x$ regimes, where, in addition,
one expects respectively $\Delta q \sim\Delta\bar{q}$ and 
$\Delta q \gg \Delta\bar{q}$. Then, {\tt NNPDF} parton sets include all
the experimental information relevant for this study.

Concerning the procedure used for PDF determination, {\tt NNPDF} parton sets
are based on a methodology which uses a Monte Carlo sampling and representation
of PDFs, and a parameterization of PDFs based on neural networks 
with a redundant number of free parameters. Both these features allow for
providing a PDF set in which the {\it procedural} 
uncertainty (due to the methodology used to determine PDFs from data) 
is reduced as much as possible. Most importantly, thanks to the 
neural network parameterization, the PDF behavior 
at small and large $x$ can deviate from the 
powerlike functional form usually assumed in other PDF parameterizations. 
All {\tt NNPDF} parton sets, 
both unpolarized and polarized, are determined within this methodology in 
a mutually consistent way.  

}

\paragraph{Small-$x$ behavior}
{
What the behavior of polarized PDFs should be at $x\to 0$ is presently 
not well understood. 
Nevertheless, several models attempt to provide an estimate 
of the polarized, neutral-current, virtual-photon, DIS
structure function $g_1$ at small-$x$ values. 
Arguments based on the dominance of known Regge poles~\citep{Heimann:1973hq}
lead to the expectation 
\begin{equation}
g_1(x)\xrightarrow{x\sim 0} x^{-\lambda}
\mbox{\, ,}
\label{eq:exp1}
\end{equation}
where $\lambda$ is the intercept of the $a_1(1260)$ meson Regge trajectory in 
the isovector channel and the $f_1(1285)$ meson trajectory in the isoscalar
channel. Roughly, this leads to~\citep{Bass:2006dq}
\begin{equation}
-0.4\leq\lambda_{a_1} \approx \lambda_{f_1}\leq -0.18
\mbox{\, .}
\label{eq:exp2}
\end{equation}
A model of the pomeron based on nonperturbative gluon 
exchange~\citep{Bass:1994xb} gives the singular behavior
\begin{equation}
g_1(x)\xrightarrow{x\sim 0} A (-2\ln x - 1)
\mbox{\, ,}
\label{eq:exp3}
\end{equation}
while it has also been argued~\citep{Close:1994he} that 
it is possible to induce the extremely singular behavior
\begin{equation}
g_1(x)\xrightarrow{x\sim 0} \frac{B}{x\ln^{2}x}
\mbox{\, ,}
\label{eq:exp4}
\end{equation}
where $A$ and $B$ are normalization coefficients determined from a fit to
experimental data. 

Regge theory is expected to be valid only at low $Q^2$ and a  
behavior of the form~(\ref{eq:exp1}) is unstable under DGLAP 
evolution~\citep{Ball:1995ye,Altarelli:1998nb,Gehrmann:1995ut}.
Indeed, as $Q^2$ increases, contributions proportional to $\ln\left(1/x\right)$
enter the evolution equations via the splitting functions, which, in the 
polarized case, all contain singularities~\citep{Ahmed:1975tj}.
Small-$x$ logarithms are included via DGLAP evolution 
up to NLO accuracy in global QCD analyses of polarized PDFs. 
However, polarized splitting functions are further enhanced 
at N$^n$LO by {\it double} logarithms of the form  $\alpha_s^n\ln^{2n-1} x$,
which correspond to the ladder diagrams with quark and gluon exchanges
along the ladder.
Calculations summing up these contributions, either with 
non-running~\citep{Bartels:1995iu,Bartels:1996wc} 
or running~\citep{Ermolaev:2003zx}
values of the strong coupling $\alpha_s$, found that the singlet flavor
combination of proton and neutron structure functions, $g_1^p + g_1^n$,
should diverge more rapidly than the nonsinglet combination, $g_1^p - g_1^n$,
as $x$ goes to zero, {\it i.e.}
\begin{equation}
|g_1^p + g_1^n|-|g_1^p - g_1^n| \geq 0
\mbox{\, , \,}
x\to 0
\mbox{\, .}
\label{eq:exp6}
\end{equation}
The next-to-next-to-leading order (NNLO) corrections to the polarized
splitting functions have been computed very recently~\citep{Moch:2014sna}:
these are found to be small and unproblematic down to at least 
$x\sim 10^{-4}$.

Finally, coherence arguments~\citep{Brodsky:1989db,Brodsky:1994kg}
suggest that, at a typical nucleon scale, the polarized gluon distribution 
$\Delta g(x)$ should be related to its unpolarized counterpart, $g(x)$, 
according to
\begin{equation}
\frac{\Delta g(x)}{g(x)}\xrightarrow{x\sim 0} 2x
\mbox{\, .}
\label{eq:exp5}
\end{equation}

In order to test the validity of expectations~(\ref{eq:exp1}-\ref{eq:exp5}),
corresponding predictions are made using 
{\tt NNPDFpol1.1}~\citep{Nocera:2014gqa} and {\tt NNPDF2.3}~\citep{Ball:2012cx} 
parton sets. No model assumptions were imposed for constraining the small-$x$ 
behavior of these PDFs, except the requirement that polarized PDFs must be 
integrable, {\it i.e.} they have finite first moments. 

In order to study the potential impact of future measurements at an EIC, 
the {\tt NNPDFpolEIC-B}~\citep{Ball:2013tyh} parton set will be used too. 
This was determined from a fit to the inclusive DIS data in {\tt NNPDFpol1.1}, 
supplemented with simulated inclusive DIS pseudodata at a future EIC
down to $x\sim 10^{-5}$.
These pseudodata were generated assuming that the {\it true} underlying set
of PDFs is that in Ref.~\citep{deFlorian:2009vb}, even though the behavior
of polarized PDFs at such small-$x$ values is not known.
Hence, the {\tt NNPDFpolEIC-B} parton set does encode information on 
the potential reduction of PDF uncertainties from future DIS measurements 
at an EIC, but definitely does not encode additional information 
on the small-$x$ behavior of polarized PDFs.

The impact of future measurements at an EIC was previously addressed also in 
Ref.~\citep{Aschenauer:2012ve}, where projected neutral-current  
DIS and SIDIS artificial data were added to the {\tt DSSV} 
polarized PDF set of Ref.~\citep{deFlorian:2009vb}. 
In comparison to the {\tt NNPDFpolEIC-B}
determination, pseudodata were generated assuming the same underlying set
of PDFs, but they were then included in a global QCD analysis using a 
substantially different fitting methodology. For this reason, since the
{\tt NNPDF}~\citep{Ball:2013tyh} and the {\tt DSSV}~\citep{Aschenauer:2012ve}
studies found similar PDF uncertainties, error bands can be reasonably
trusted in the {\tt NNPDFpolEIC-B} parton set.  
Only real data will determine what should be the central value instead.
 
In Fig.~\ref{fig:smallg1} (left panel), the spin-dependent structure 
function of the proton, $g_1$, is shown as a function of 
$x$ at $Q^2=4$ GeV$^2$, together with 
available experimental data from SMC~\citep{Adeva:1998vv},
E143~\citep{Abe:1998wq}, COMPASS~\citep{Alekseev:2010hc} and 
HERMES~\citep{Airapetian:2006vy} experiments, and 
model expectations, Eqs.~(\ref{eq:exp1}-\ref{eq:exp4}).
In Fig.~\ref{fig:smallg1} (right panel), the quantity defined in
Eq.~(\ref{eq:exp6}) is displayed at $Q^2=4$ GeV$^2$.
In Fig.~\ref{fig:PDFg} (left panel), the ratio of polarized 
to unpolarized gluon PDFs, $\Delta g/g$,
is plotted in the small-$x$ region at $Q^2=4$ GeV$^2$.
In Fig.~\ref{fig:alpha}, the small-$x$ effective exponents 
\begin{equation}
\alpha_q(x,Q^2) = -\frac{\partial\ln \left| q(x,Q^2) \right|}{\partial\ln x}
\label{eq:alphadef}
\end{equation}
are displayed as a function of $x$ at $Q^2=4$ GeV$^2$ for unpolarized 
{\tt NNPDF2.3} and polarized {\tt NNPDFpol1.1} PDFs, 
$q=u,\bar{u},d,\bar{d},\bar{s},g$ and 
$q=\Delta u,\Delta\bar{u},\Delta d,\Delta\bar{d},\Delta\bar{s},\Delta g$
respectively.
The corresponding values at $x=10^{-5}$ are reported in 
Tab.~\ref{tab:prepexp-l}.
All predictions based on {\tt NNPDF} parton sets are obtained at NLO
QCD accuracy, and corresponding uncertainties are nominal one-$\sigma$ bands. 
\begin{figure*}[t]
 \centering
 \includegraphics[scale=0.4]{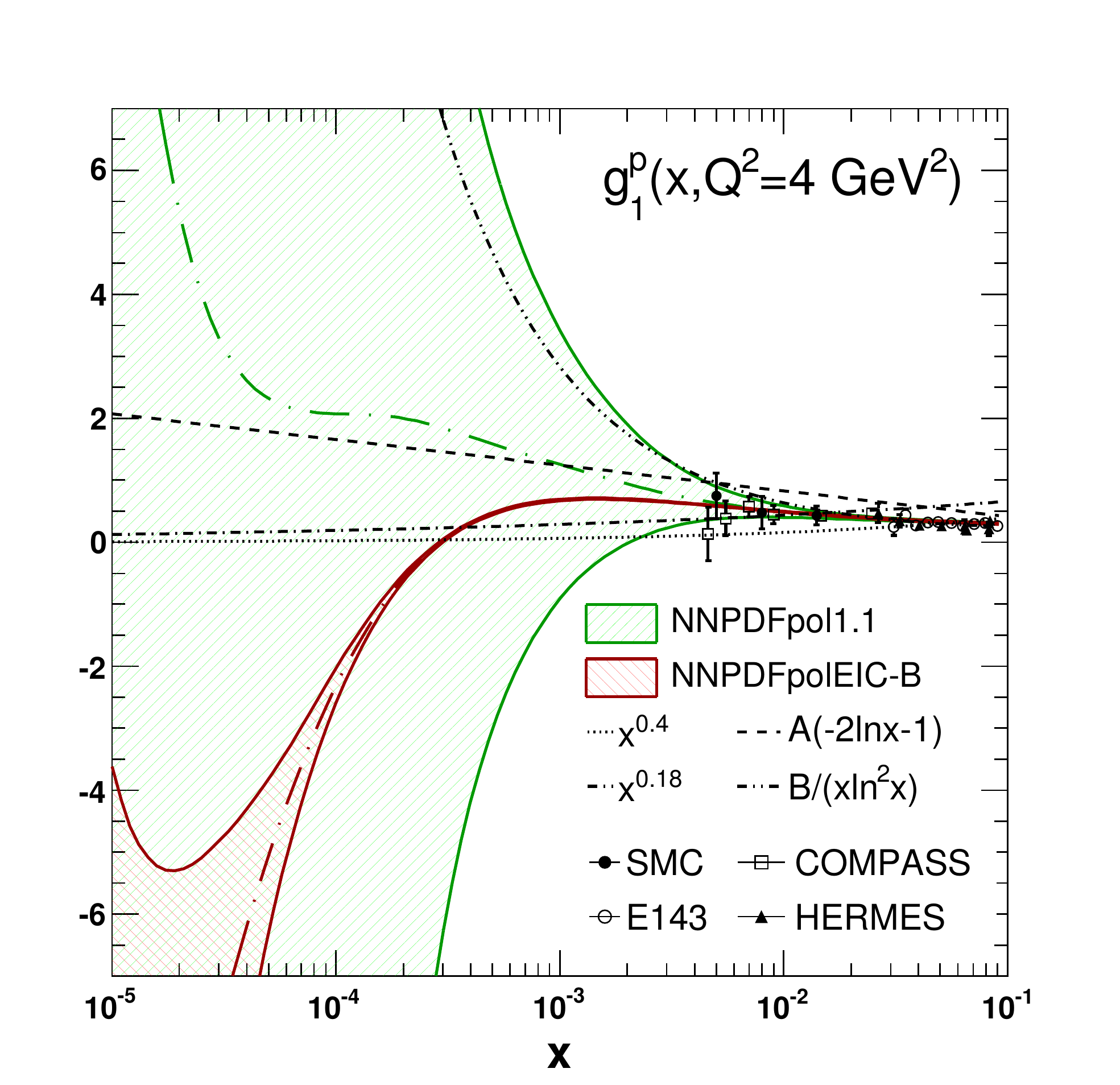}
 \includegraphics[scale=0.4]{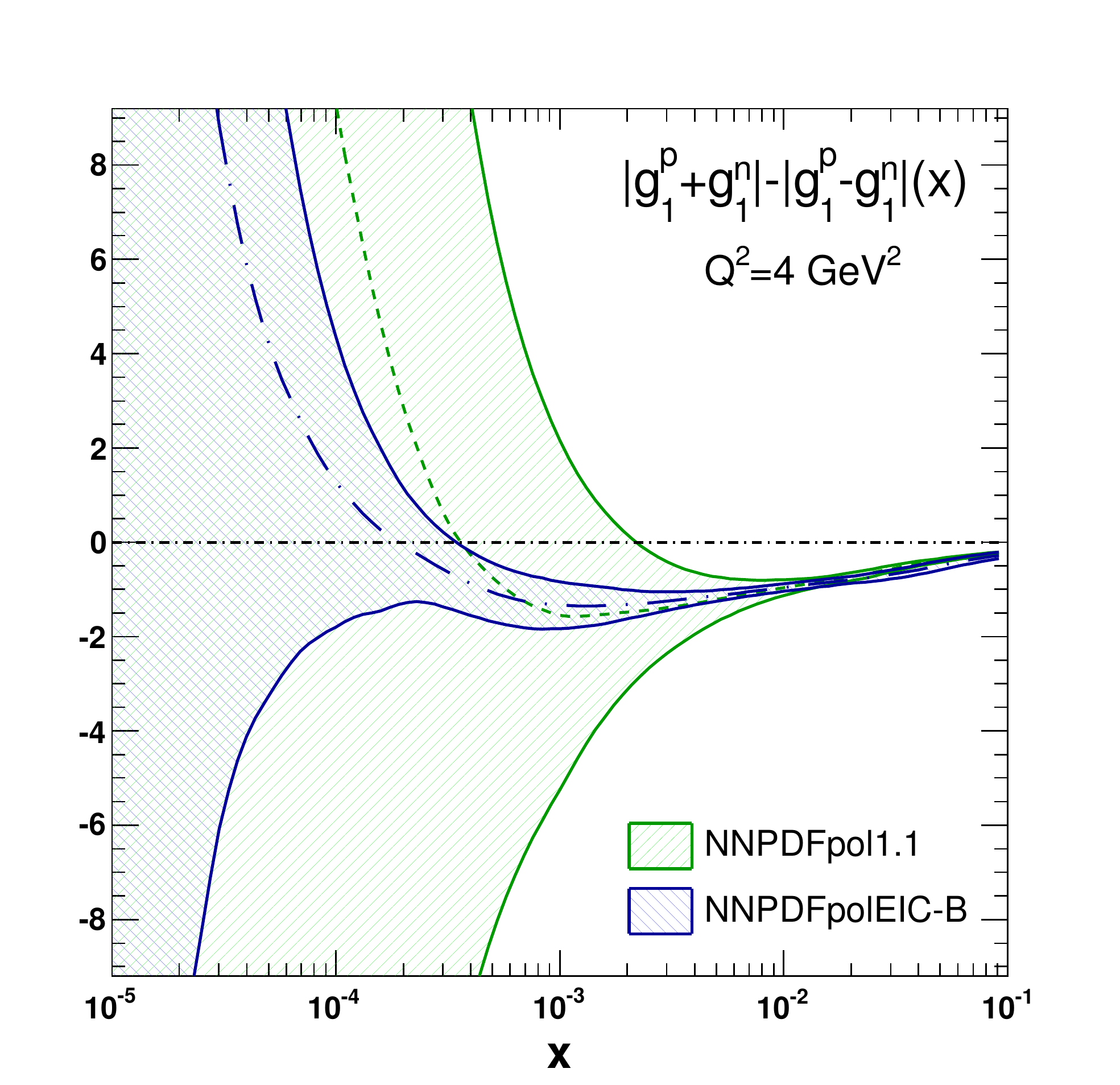}\\
 \caption{\small (Left panel). The spin-dependent structure function 
of the proton, $g_1^p$, as a function of $x$ at $Q^2=4$ GeV$^2$.
Predictions are obtained using PDFs from 
{\tt NNPDFpol1.1}~\citep{Nocera:2014gqa}
and {\tt NNPDFpolEIC-B}~\citep{Ball:2013tyh} parton sets. 
Experimental data at small $x$ from SMC~\citep{Adeva:1998vv},
E143~\citep{Abe:1998wq}, COMPASS~\citep{Alekseev:2010hc} and 
HERMES~\citep{Airapetian:2006vy}, and theoretical expectations, 
Eqs.~(\ref{eq:exp1}-\ref{eq:exp4}), are also shown.
The values of the normalization coefficients $A$ and $B$ entering 
predictions~(\ref{eq:exp2})-(\ref{eq:exp3}) are taken respectively from 
Ref.~\citep{Bass:1994xb} and Ref.~\citep{Close:1994he}:
$A=0.09$ and $B=0.135$.
(Right panel). The prediction, Eq.~(\ref{eq:exp6}), at $Q^2=4$ GeV$^2$
obtained using the same PDF sets as in left panel. Note that 
Eq.~(\ref{eq:exp6}) is fulfilled whenever curves are positive.}
 \label{fig:smallg1}
\end{figure*}
\begin{figure*}[t]
 \centering
 \includegraphics[scale=0.4]{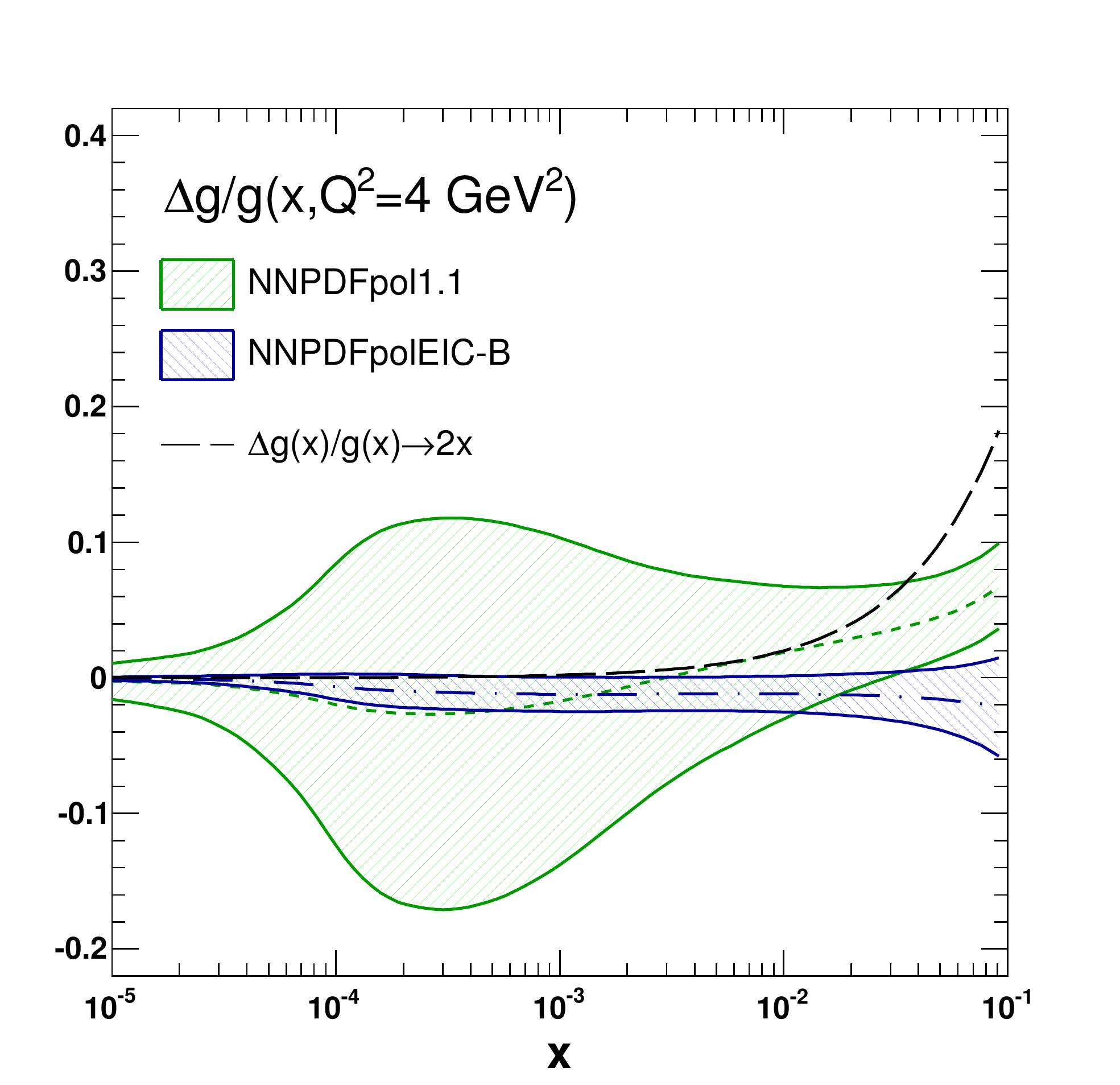}
 \includegraphics[scale=0.4]{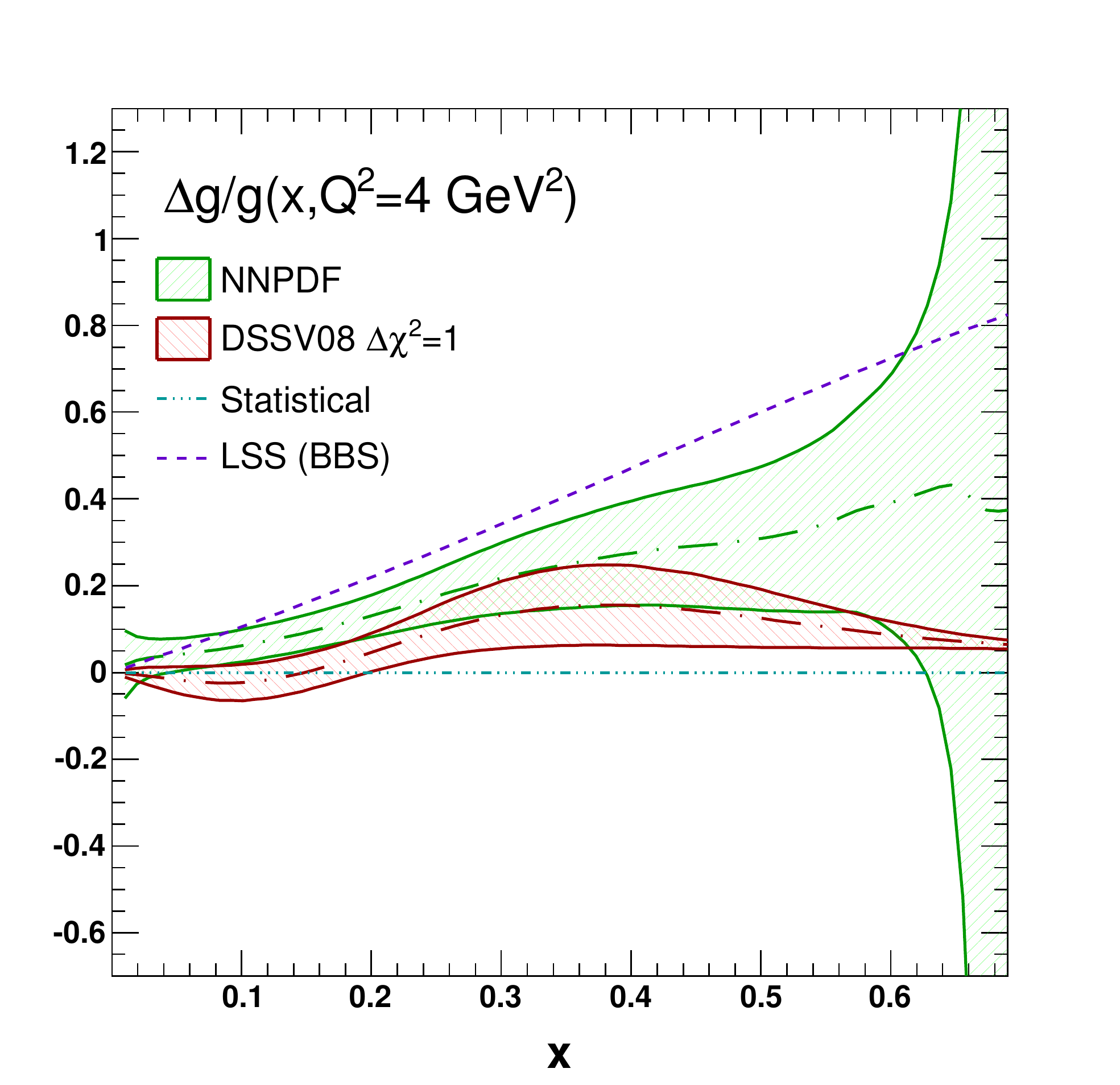}\\
 \caption{\small The ratio of polarized to unpolarized gluon 
PDFs, $\Delta g/g$, as a function of $x$ at $Q^2=4$ GeV$^2$ both in the
small- (left panel) and large-$x$ (right panel) regions. Predictions are 
obtained using NLO polarized {\tt NNPDFpol1.1}~\citep{Nocera:2014gqa} 
and unpolarized {\tt NNPDF2.3}~\citep{Ball:2012cx} parton sets.
Predictions obtained using polarized 
{\tt NNPDFpolEIC-B}~\citep{Ball:2013tyh} parton set 
instead of {\tt NNPDFpol1.1} are also displayed 
in the small-$x$ region. 
Expectations from both Eq.~(\ref{eq:exp5}) and 
statistical~\citep{Bourrely:2001du} and LSS(BBS)~\citep{Leader:1997kw} 
parameterizations, available in the small- 
and large-$x$ regions respectively, are shown for comparison.}
 \label{fig:PDFg}
\end{figure*}
\begin{figure*}[t]
\centering
\includegraphics[scale=0.85]{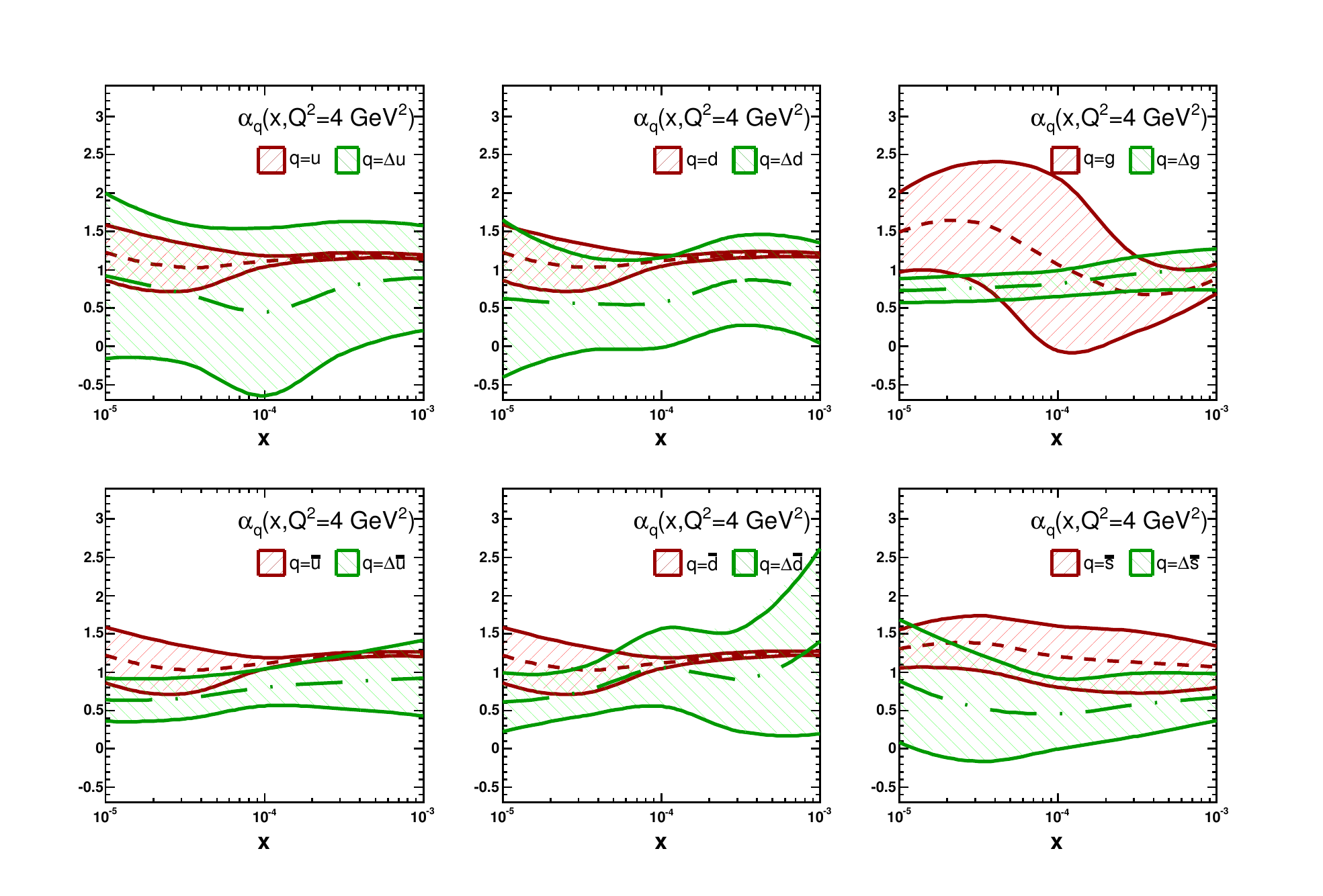}
\caption{\small Small-$x$ effective exponents, Eq.~(\ref{eq:alphadef}), 
at $Q^2=4$ GeV$^2$ as a function of $x$.}
\label{fig:alpha}
\end{figure*}
\begin{table*}[t]
 \centering
 \footnotesize
 \begin{tabular}{lccccccc}
  \toprule
  PDF set                           & Ref.        
  & $u$        & $\bar{u}$      & $d$        & $\bar{d}$      & $\bar{s}$      & $g$           \\
  \midrule
  {\tt NNPDF2.3}                    & ~\citep{Ball:2012cx}   
  & $1.22\pm 0.36$ & $1.23\pm 0.36$ & $1.22\pm 0.36$ & $1.23\pm 0.36$ & $1.31\pm 0.25$ & $1.49\pm 0.52$ \\
  {\tt NNPDFpol1.1}                 &~\citep{Nocera:2014gqa}   
  & $0.92\pm 1.08$ & $0.61\pm 0.39$ & $0.63\pm 1.03$ & $0.64\pm 0.28$ & $0.81\pm 0.80$ & $0.73\pm 0.15$ \\
  \bottomrule
 \end{tabular}
\caption{\small The values of small-$x$ effective exponents, Eq.~(\ref{eq:alphadef}),
at $x=10^{-5}$ and $Q^2=4$ GeV$^2$.}
\label{tab:prepexp-l}
\end{table*}

Inspection of Figs.~\ref{fig:smallg1}-\ref{fig:PDFg}-\ref{fig:alpha} and 
Tab.~\ref{tab:prepexp-l} allows for drawing the following conclusions.
\begin{itemize}
 \item Because of the lack of experimental information, the prediction
for the small-$x$ behavior of the proton structure $g_1^p$ obtained
from the {\tt NNPDFpol1.1} parton set is largely uncertain. As a consequence,
it does not allow for discriminating between powerlike Regge expectations, 
Eqs.~(\ref{eq:exp1}-\ref{eq:exp2}), and other behaviors, 
Eqs.~(\ref{eq:exp3}-\ref{eq:exp4}).
A substantial reduction of the uncertainty on $g_1^p$,
up to one order of magnitude,
is expected to be provided by a future EIC, 
as suggested by the corresponding prediction obtained
using the {\tt NNPDFpolEIC-B} parton set. In this case, one will be able to
discriminate between expectations~(\ref{eq:exp1}-\ref{eq:exp6}).

 \item The {\tt NNPDFpol1.1} prediction does not support the
expectation~(\ref{eq:exp6}) at moderately small-$x$ values, 
$x\gtrsim 10^{-3}$. This conclusion
is consistent with the results reported by the SMC 
experiment on singlet and nonsinglet structure function combinations, 
$g_1^p+g_1^n$ and $g_1^p-g_1^n$~\citep{Adeva:1998vv}. 
It was suggested in Ref.~\citep{Blumlein:1996hb} that the discrepancy 
observed between Eq.~(\ref{eq:exp6}) and SMC data (and hence the 
{\tt NNPDFpol1.1} prediction) may be explained
by significant corrections due to sub-leading terms,
which may eventually alter Eq.~(\ref{eq:exp6}).
If anything, the 
expectation~(\ref{eq:exp6}) is fulfilled by the corresponding prediction
obtained with {\tt NNPDFpol1.1} at $x\lesssim 10^{-3}$. 
However, the latter may not be reliable, because
of the complete lack of data. Also, all double-logarithmic contributions
are neglected in the DGLAP evolution of the structure function $g_1$: these
contributions may become important in the small-$x$ region
and may be resummed to all orders in $\alpha_s$~\citep{Ermolaev:2003zx}.

 \item In order to get a feeling of the potential reduction of the uncertainty
attained at a future EIC, the prediction for Eq.~(\ref{eq:exp6})
obtained using the {\tt NNPDFpolEIC-B} parton set is also shown in the 
right panel of Fig.~\ref{fig:smallg1}. 
However, in the {\tt NNPDFpolEIC-B} fit only
pseudodata for an EIC with a proton beam were included; in order to 
properly address expectation~(\ref{eq:exp6}), 
it would be necessary to measure the neutron structure function
$g_1^n$ in addition. This could be done with a beam of polarized $^3$He.
Such a measurement will also be required
to improve the accuracy with which the Bjorken sum rule~\citep{Bjorken:1966jh} 
can be checked in the deeply small-$x$ region. 
Indeed, in Ref.~\citep{Ball:2013lla} it was pointed out that a 
largely uncertain, and potentially substantial,
contribution to it may arise in this region.

 \item The expectation for the ratio of polarized to unpolarized
gluon distribution, Eq.~(\ref{eq:exp5}), is consistent with the
prediction obtained using the {\tt NNPDFpol1.1} parton set. However, 
this quantity remains largely uncertain at
$x\lesssim 4\cdot 10^{-3}$, because of the lack of experimental data
in this region. The potential reduction of this uncertainty,
due to a future EIC, can be appreciated 
by comparing the predictions obtained using either the {\tt NNPDFpol1.1}
or the {\tt NNPDFpolEIC-B} parton sets. At $x\sim 10^{-4}$, the uncertainty 
estimate from the latter may be smaller than that from the former 
up to one order of magnitude.

\item The effective exponents $\alpha_q$, Eq.~(\ref{eq:alphadef}), 
estimate the powerlike
behavior of PDFs at sufficiently small-$x$ values, where 
the latter can be approximated as $q\sim x^{-\alpha_q}$; $q$
denotes either unpolarized or polarized distributions,
respectively $q=u$, $\bar{u}$, $d,\bar{d}$, $\bar{s}$, $g$ or 
$q=\Delta u$, $\Delta\bar{u}$, $\Delta d$, $\Delta\bar{d}$, 
$\Delta\bar{s}$, $\Delta g$.
Note that the definition~(\ref{eq:alphadef}) differs from that 
given by Eq.~(67) in Ref.~\citep{Ball:2013lla}, in that the derivative 
with respect to $x$ of both the numerator and the denominator is used.
In comparison to the latter definition, subasymptotic corrections, 
which may become 
negligible only at extremely small values of $x$ (and hence may contribute 
to the effective exponents  
significantly), are taken into account in the former definition.
For this reason Eq.~(\ref{eq:alphadef}) is used in the present study.
Similar arguments also hold at large $x$, hence the definition 
of the asymptotic exponents given by Eq.(68) in Ref.~\citep{Ball:2013lla} 
will be consistently 
supplemented with the derivative of both the numerator and the denominator,
see Eq.~(\ref{eq:betadef}) below.

\item Results in Tab.~\ref{tab:prepexp-l} and in 
Fig.~\ref{fig:alpha} show that the behavior of effective exponents for 
unpolarized PDFs is almost insensitive to the quark/antiquark flavor: indeed,
the effective exponents for $u$, $\bar{u}$, $d$, $\bar{d}$ distributions are
very similar among each others. This conclusion does not hold for
polarized PDFs, whose effective exponents show a less regular behavior 
than their unpolarized counterparts. 

\item At sufficiently small values of $x$, the effective exponents 
are expected to enter an asymptotic regime, {\it i.e.} to become constant. 
Such a behavior is clearly visible for the polarized gluon distribution at 
$x\lesssim 10^{-4}$. For quark and antiquark PDFs, the effective exponents 
are still slightly varying at $x\sim 10^{-5}$, thus suggesting that 
the asymptotic regime is reached at smaller values of $x$.
In the polarized case, experimental data used to constrain PDFs are located
at $x\gtrsim 10^{-3}$: because below this value the stability of the 
effective exponents is not reached, one should conclude that the 
effect of data on extrapolation is rather mild.

\end{itemize}

}

\paragraph{Large-$x$ behavior}
{
The behavior of polarized PDFs at $x\to 1$ is predicted by a number of
different theoretical models. To first approximation, the constituent quarks 
in the nucleon are described by SU(6) wave functions with zero orbital angular
momentum~\citep{Close:1974ux}. In fact, SU(6) symmetry is known to be 
broken~\citep{Close:1973xw} and, depending on the details of SU(6)-breaking
mechanisms, different behaviors of valence quarks may arise.
For instance, the Relativistic Constituent Quark Model (RCQM) 
assumes that SU(6) symmetry is broken via a color hyperfine interaction
between quarks~\citep{Isgur:1998yb}. This leads to a non-zero quark 
orbital angular momentum, and a consequent reduction of the valence
quark contributions to the nucleon spin at large $x$. 
Different mechanisms of SU(6) breaking, consistent with duality, were also
included in Quark-Hadron Duality (QHD) models~\citep{Close:2003wz}. 

Statistical models also predict the behavior of polarized PDFs at large $x$.
They treat the nucleon as a gas of massless partons at thermal equilibrium,
using both chirality and DIS data to constrain the thermodynamic potential of 
each parton species~\citep{Bourrely:2001du}.
Both scalar and axial diquark channels have been included in a modified
Nambu-Jona-Lasinio (NJL) model~\citep{Cloet:2005pp}.

An approach to nucleon structure based on Dyson-Schwinger Equations (DSE), 
in which it
is described according to the relevant Poincar\'{e}-covariant Faddeev equation,
has been studied recently~\citep{Roberts:2013mja}, assuming the simplification 
that the sum of soft, dynamic, non-pointlike diquark correlations 
approximates the quark-quark scattering matrix.
Further expectations are provided by chiral 
soliton~\citep{Wakamatsu:2014asa},
instanton~\citep{Kochelev:1997ux} and 
bag~\citep{Boros:1999tb} models.

In leading-order (LO) perturbative QCD (pQCD), 
at large $x$ and large $Q^2$, the 
valence quark orbital angular momentum may be assumed to be negligible,
thus leading to hadron helicity conservation~\citep{Farrar:1975yb}. 
Parameterizations of the world DIS data have been made with and without this
assumption. Specifically, this is included in the LSS(BBS) parton 
determination~\citep{Leader:1997kw}, while Fock states with nonzero quark
orbital angular momentum are considered in a parameterization by Avakian
{\it et al.}~\citep{Avakian:2007xa}.

In Tab.~\ref{tab:models} model expectations for various ratios of
polarized/unpolarized PDFs and spin-dependent neutron and proton virtual
photoabsorption asymmetries, $A_1^n$ and $A_1^p$, 
at $x\to 1$ are collected. 
Specifically, results for SU(6)~\citep{Close:1974ux},
RCQM~\citep{Isgur:1998yb}, QHD with two different SU(6)-breaking 
mechanisms~\citep{Close:2003wz}, NJL~\citep{Cloet:2005pp}, DSE
{\it realistic} and {\it contact}~\citep{Roberts:2013mja} models, and the
LO pQCD prediction assuming zero orbital angular momentum~\citep{Farrar:1975yb}
are reported. The corresponding {\tt NNPDF} predictions are also shown 
at $Q^2=4$ GeV$^2$ for different values of $x$.
\begin{table*}[t]
 \centering
 \footnotesize
 \begin{tabular}{lccccccc}
  \toprule
  Model & Refs. & 
  $d/u$          & $\Delta d/\Delta u$ & $\Delta u/u$   & $\Delta d/d$   & $A_1^n$        & $A_1^p$ \\
  \midrule
  SU(6)          & ~\citep{Close:1974ux}                                  &
  $1/2$          & $-1/4$              & $2/3$          & $-1/3$         & $0$            & $5/9$   \\
  RCQM           & ~\citep{Isgur:1998yb}                                  &
  $0$            & $0$                 & $1$            & $-1/3$         & $1$            & $1$     \\
  QHD ($\sigma_{1/2}$)          & ~\citep{Close:2003wz}                                   &
  $1/5$          & $1/5$               & $1$            & $1$            & $1$            & $1$     \\
  QHD ($\psi_{\rho}$)          & ~\citep{Close:2003wz}                                   &
  $0$            & $0$                 & $1$            & $-1/3$         & $1$            & $1$     \\
  NJL            & ~\citep{Cloet:2005pp}                                  &
  $0.20$         & $-0.06$             & $0.80$         & $-0.25$        & $0.35$         & $0.77$  \\
  DSE ({\it realistic})          & ~\citep{Roberts:2013mja}                  &
  $0.28$         & $-0.11$             & $0.65$         & $-0.26$        & $0.17$         & $0.59$  \\
  DSE ({\it contact})          & ~\citep{Roberts:2013mja}                &
  $0.18$         & $-0.07$             & $0.88$         & $-0.33$        & $0.34$         & $0.88$  \\
  pQCD           & ~\citep{Farrar:1975yb}                  &
  $1/5$          & $1/5$               & $1$            & $1$            & $1$            & $1$     \\
  \midrule
  {\tt NNPDF} ($x=0.7$) &~\citep{Ball:2012cx,Nocera:2014gqa}                     & 
  $0.22\pm 0.04$ & $-0.07\pm 0.12$      & $0.07\pm 0.05$ & $-0.19\pm 0.34$ & $0.41\pm 0.31$ & $0.75\pm 0.07$ \\
  {\tt NNPDF} ($x=0.8$) &~\citep{Ball:2012cx,Nocera:2014gqa}                     & 
  $0.18\pm 0.09$ & $\ \ 0.12\pm 0.23$      & $0.70\pm 0.13$ & $\ \ 0.34\pm 0.67$ & $0.57\pm 0.61$ & $0.75\pm 0.12$ \\
  {\tt NNPDF} ($x=0.9$) &~\citep{Ball:2012cx,Nocera:2014gqa}                     & 
  $0.06\pm 0.49$ & $\ \ 0.51\pm 0.69$      & $0.61\pm 0.48$ & $\ \ 0.85\pm 6.55$ & $0.36\pm 0.61$ & $0.74\pm 0.34$ \\
  \bottomrule
  \end{tabular}
\caption{\small A collection of several model expectations for various ratios of
polarized/unpolarized PDFs and spin-dependent neutron and proton asymmetries,
$A_1^n$ and $A_1^p$, at $x\to 0$. 
The {\tt NNPDF} prediction, obtained using unpolarized 
{\tt NNPDF2.3}~\citep{Ball:2012cx}
and polarized {\tt NNPDFpol1.1}~\citep{Nocera:2014gqa} parton sets, 
is shown at $Q^2=4$ GeV$^2$ for different values of $x$.}
\label{tab:models}
\end{table*}

In Fig.~\ref{fig:PDFg} (right panel) and Fig.~\ref{fig:PDFq} respectively, 
the ratios of polarized to unpolarized gluon distributions, $\Delta g/g$,
and total $u$ and $d$ quark combinations, 
$\Delta q^+/q^+=(\Delta q+\Delta\bar{q})/(q+\bar{q})$, $q=u,d$,
are displayed as a function of $x$ at $Q^2=4$ GeV$^2$. 
Expectations from statistical~\citep{Bourrely:2001du},
NJL~\citep{Cloet:2005pp} and QHD~\citep{Close:2003wz} 
(with two different SU(6)-breaking 
mechanisms) models and from LSS(BBS)~\citep{Leader:1997kw} 
and Avakian {\it et al.}~\citep{Avakian:2007xa} parameterizations
are shown. Notice that not all of them are available for the gluon.
The curve labeled {\tt DSSV08} 
is obtained using the polarized {\tt DSSV08}~\citep{deFlorian:2009vb}
and the unpolarized {\tt MRST}~\citep{Martin:2002aw} parton sets
(the latter was used for reference in~\citep{deFlorian:2009vb}).
The uncertainty is the Hessian uncertainty computed assuming
$\Delta\chi^2=1$. This choice may lead to somewhat underestimated 
uncertainties: it is well known that, in global fits based on Hessian
methodology, a tolerance $\Delta\chi^2=T>1$ is needed for faithful 
uncertainty estimation. Indeed, in
Ref.~\citep{deFlorian:2009vb}, uncertainty estimates obtained from 
the Lagrange multiplier method with $\Delta\chi^2/\chi^2=2\%$
(roughly corresponding to $T\sim 8$) were recommended to be more reliable. 
In this case, the uncertainties of the {\tt DSSV08} curves would be larger 
than those shown in Figs.~\ref{fig:PDFg}-\ref{fig:PDFq} by a factor $\sqrt{T}$.
\begin{figure*}[t]
 \centering
 \includegraphics[scale=0.4]{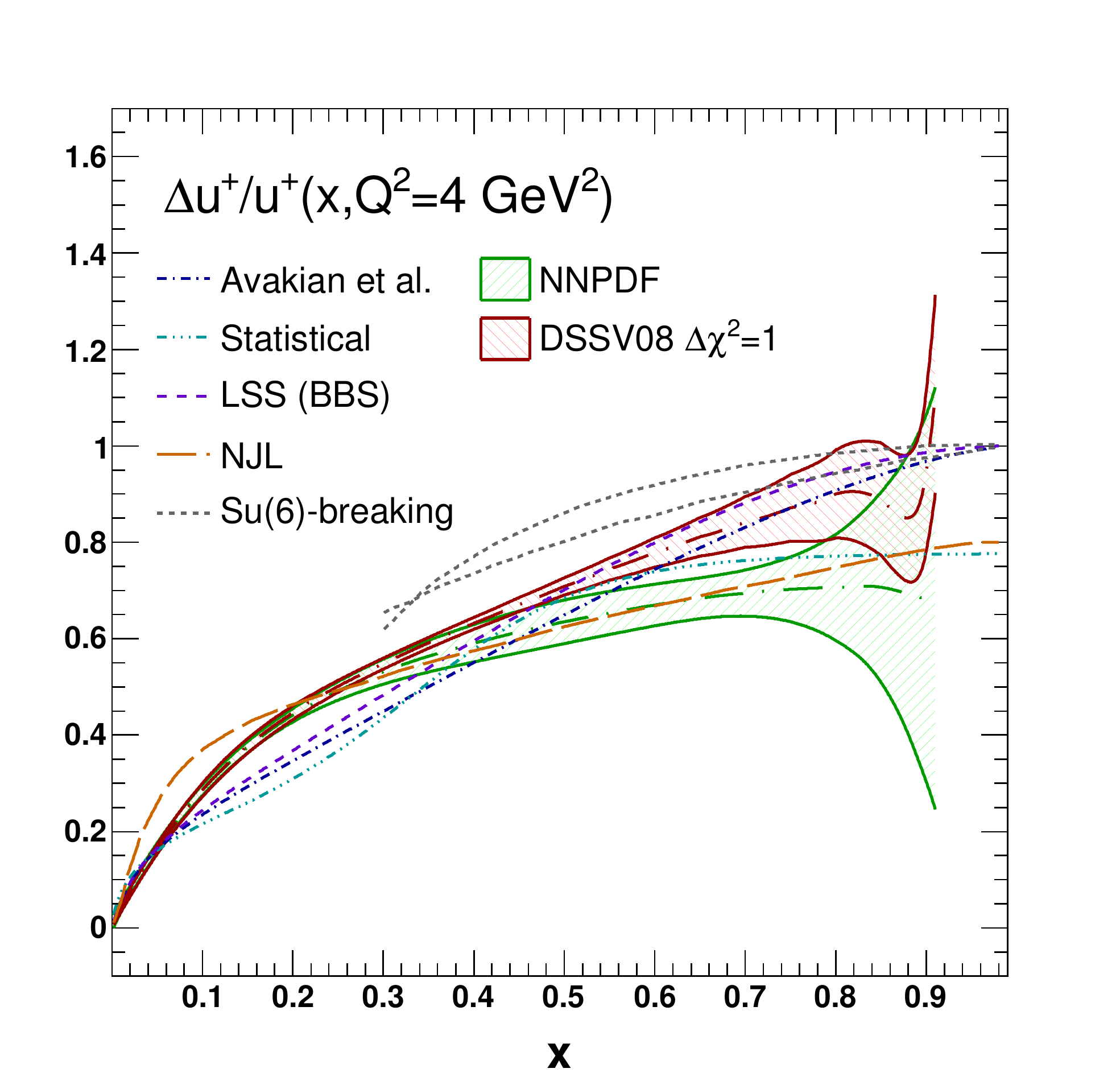}
 \includegraphics[scale=0.4]{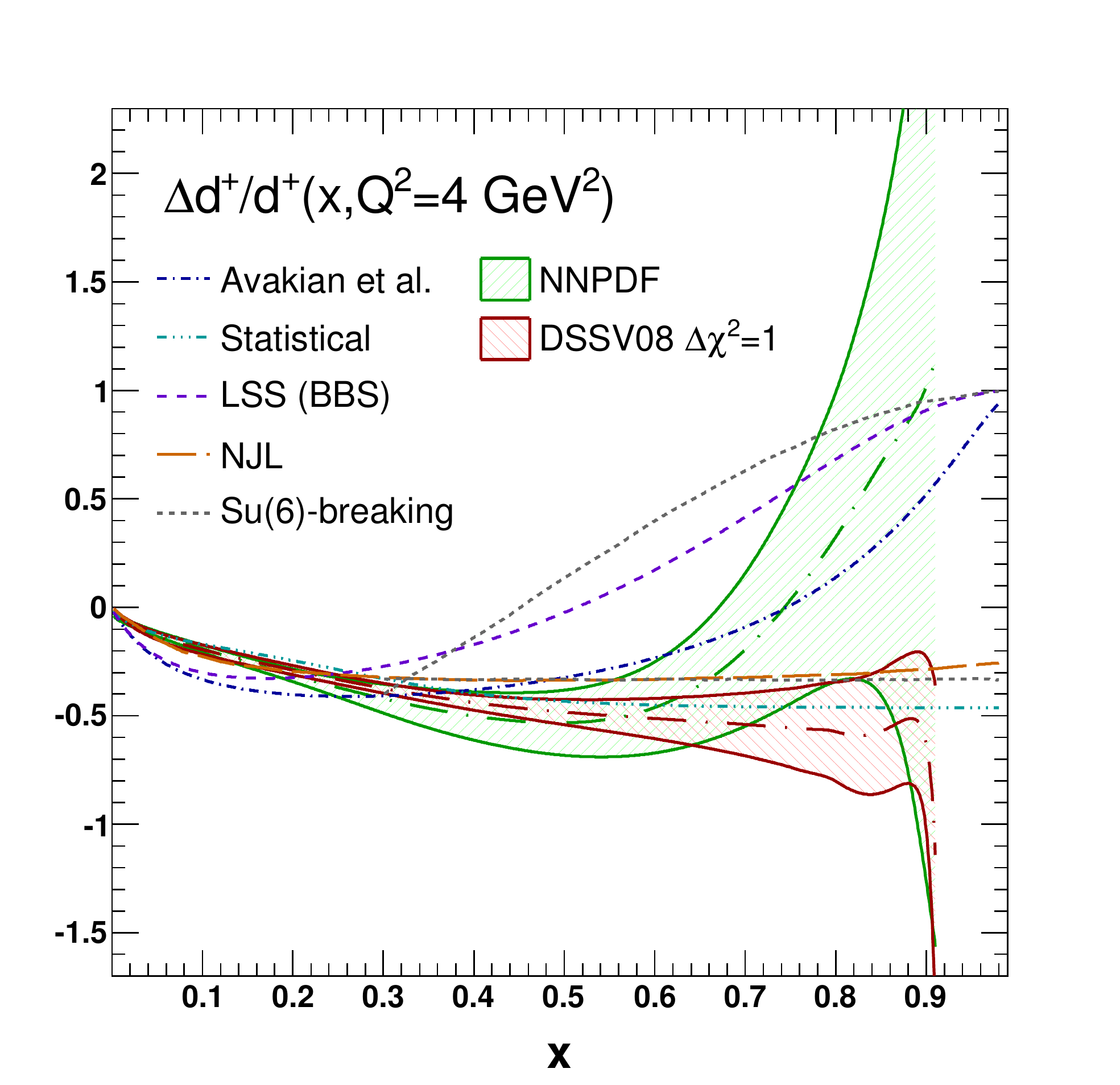}\\
 \caption{\small The ratio of polarized to unpolarized total $u$ (left) 
and $d$ (right) quark combinations as a function of $x$. 
Predictions obtained with {\tt NNPDF} and {\tt DSSV08} parton sets
are compared with expectations provided by various theoretical models,
see the text for details. All results are displayed at $Q^2=4$ GeV$^2$.}
 \label{fig:PDFq}
\end{figure*}

In Fig.~\ref{fig:asy}, the neutron and proton spin-dependent 
virtual photoabsorption asymmetries, $A_1^p$ and $A_1^n$, are displayed
as a function of $x$ at $Q^2=4$ GeV$^2$. When available,
model expectations are shown as in Fig.~\ref{fig:PDFq} and
supplemented with the RCQM prediction~\citep{Isgur:1998yb}.
\begin{figure*}[t]
 \centering
 \includegraphics[scale=0.4]{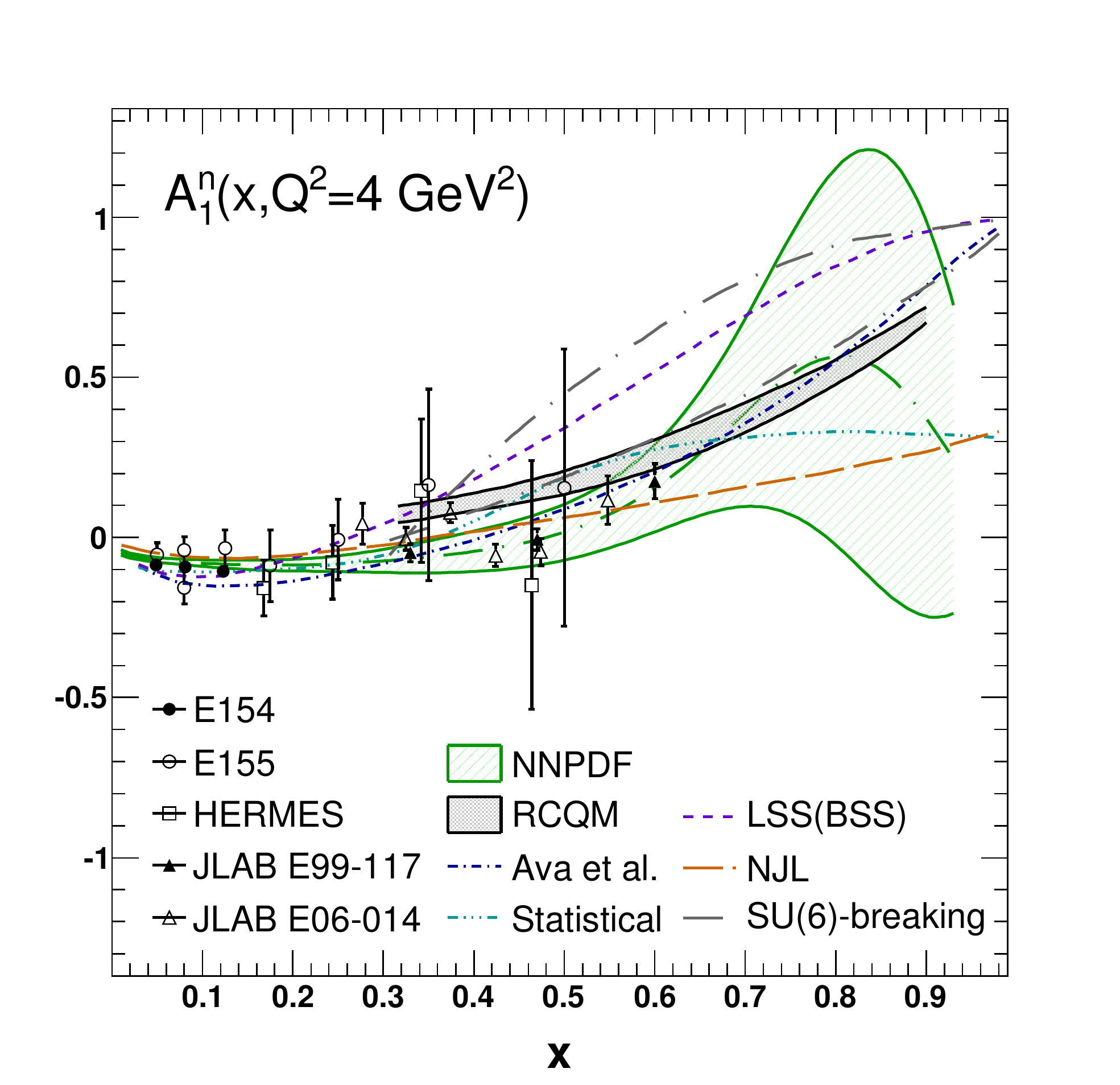}
 \includegraphics[scale=0.4]{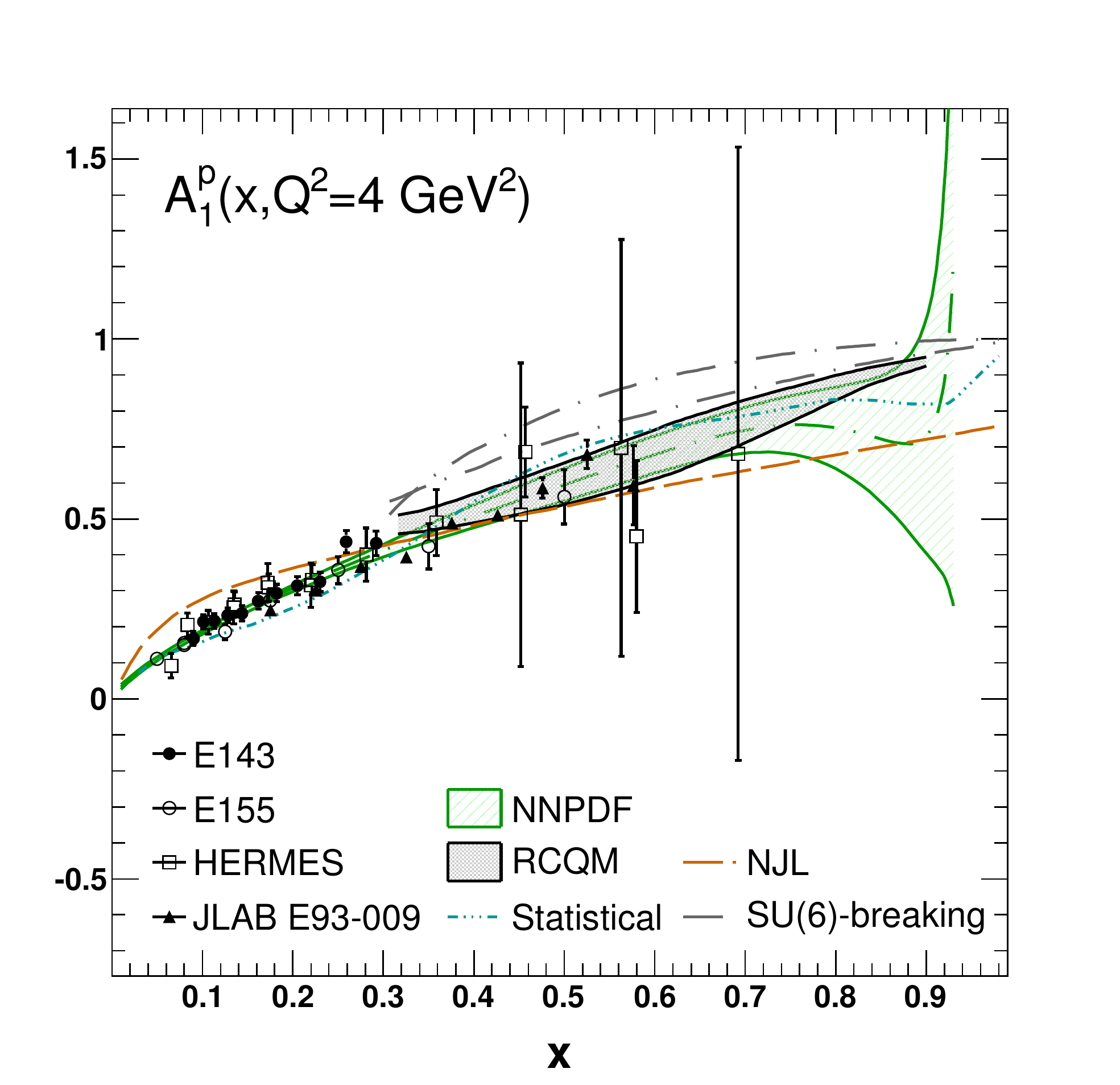}\\ 
 \caption{\small The neutron (left) and proton (right) spin-dependent 
virtual photoabsorption asymmetry, $A_1^p$ and $A_1^n$,
as a function of $x$. Predictions obtained with {\tt NNPDF} parton sets
are compared with expectations provided by various theoretical models and with
available experimental data, see the text for details. 
All results are displayed at $Q^2=4$ GeV$^2$.}
 \label{fig:asy}
\end{figure*}

In Fig.~\ref{fig:beta}, the large-$x$ effective exponents
\begin{equation}
\beta_q(x,Q^2) = \frac{\partial\ln \left| q(x,Q^2) \right|}{\partial\ln (1-x)}
\label{eq:betadef}
\end{equation}
are plotted as a function of $x$ at $Q^2=4$ GeV$^2$ 
for unpolarized and polarized PDFs, 
$q=u$, $\bar{u}$, $d$, $\bar{d}$, $\bar{s}$, $g$ and 
$q=\Delta u$, $\Delta\bar{u}$, $\Delta d$, $\Delta\bar{d}$, 
$\Delta\bar{s}$, $\Delta g$.
The corresponding values at $x=0.9$ are reported in Tab.~\ref{tab:prepexp-s}.
\begin{figure*}[t]
\centering
\includegraphics[scale=0.85]{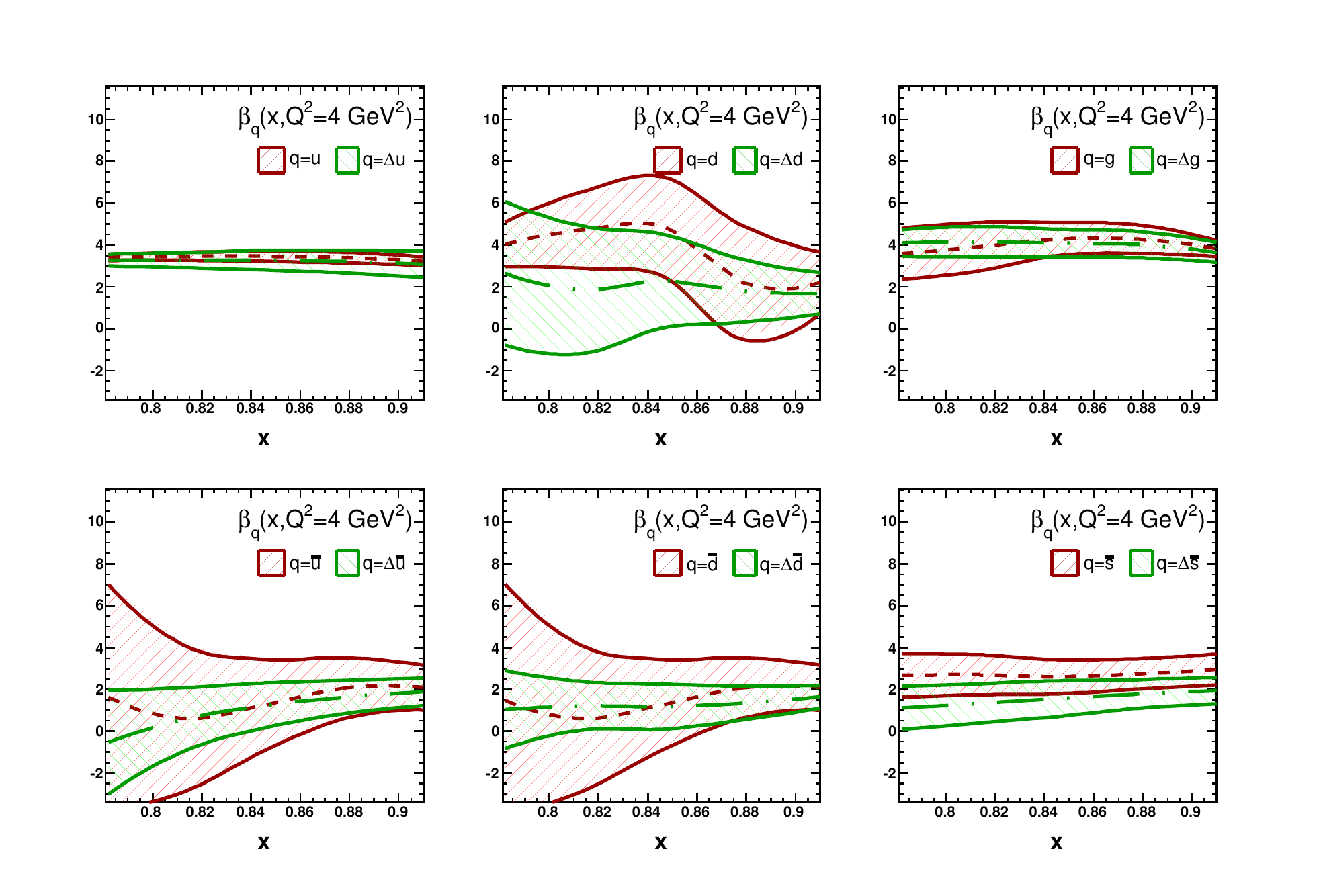}
\caption{\small Large-$x$ effective exponents, Eq.~(\ref{eq:betadef}), 
at $Q^2=4$ GeV$^2$ as a function of $x$.}
\label{fig:beta}
\end{figure*}
\begin{table*}[t]
 \centering
 \footnotesize
 \begin{tabular}{lccccccc}
  \toprule
  PDF set                           & Ref.        
  & $u$        & $\bar{u}$      & $d$        & $\bar{d}$      & $\bar{s}$      & $g$           \\
  \midrule
  {\tt NNPDF2.3}                    & ~\citep{Ball:2012cx}   
  & $3.23\pm 0.21$ & $2.09\pm 1.07$ & $2.20\pm 1.46$ & $2.09\pm 1.07$ & $2.95\pm 0.74$ & $3.82\pm 0.37$ \\
  {\tt NNPDFpol1.1}                 &~\citep{Nocera:2014gqa}   
  & $3.08\pm 0.64$ & $1.65\pm 0.55$ & $1.69\pm 0.99$ & $1.89\pm 0.66$ & $1.95\pm 0.63$ & $3.64\pm 0.47$ \\
  \bottomrule
 \end{tabular}
\caption{\small The values of large-$x$ effective exponents, 
Eq.~(\ref{eq:betadef}), at $x=0.9$ and $Q^2=4$ GeV$^2$.}
\label{tab:prepexp-s}
\end{table*}

The {\tt NNPDF} expectations, displayed 
in Tabs.~\ref{tab:models}-\ref{tab:prepexp-s} and 
in Figs.~\ref{fig:PDFg},~\ref{fig:PDFq}-\ref{fig:beta}, 
are obtained
using polarized {\tt NNPDFpol1.1}~\citep{Nocera:2014gqa} and
unpolarized {\tt NNPDF2.3}~\citep{Ball:2012cx} parton sets at NLO accuracy. 
All uncertainties are nominal one-$\sigma$ bands. 
The neutron and proton spin-dependent virtual photoabsorption asymmetries,
$A_1^n$ and $A_1^p$ in Tab.~\ref{tab:models} and Fig.~\ref{fig:asy},
are computed from the corresponding polarized and 
unpolarized structure functions $g_1$, $g_2$ and $F_1$, by inversion of 
{\it e.g.} Eq.~(18) in Ref.~\citep{Ball:2013lla}. 
Nucleon mass effects are taken into account in the relation 
between asymmetries and structure functions, consistently with the way
kinematic higher-twist terms, {\it i.e.} target mass corrections,
were included in the relation between structure functions and PDFs
when the latter were originally fitted to data~\citep{Ball:2013lla}.
Specifically, the twist-two contribution to the $g_2$ structure function 
is related to $g_1$ via Wandzura-Wilczek relation~\citep{Wandzura:1977qf},
and zero twist-three contribution to $g_2$ is assumed. 
Dynamic higher-twist contributions to the structure functions $g_1$ and $g_2$
from Wilson expansion are systematically neglected for consistency with
Refs.~\citep{Ball:2013lla,Nocera:2014gqa}. Indeed, these were shown
to be negligible in Ref.~\citep{Ball:2013lla}, though not in the large-$x$ and
small-$Q^2$ kinematic region (see also the discussion at the end of 
this section).

No model assumptions were imposed for the large-$x$
behavior of PDFs in all {\tt NNPDF} fits, except mutually consistent 
constraints from positivity of cross sections. In particular, 
the LO positivity bound was used in the polarized case,
see Eqs.~(60)-(61) in Ref.~\citep{Ball:2013lla} and discussion therein.

Inspection of Tab.~\ref{tab:models} and 
Figs.~\ref{fig:PDFg},~\ref{fig:PDFq}-\ref{fig:beta} allows for drawing the 
following conclusions.
\begin{itemize}

 \item A comparison between {\tt NNPDF} predictions and 
model expectations for PDF ratios and asymmetries allows for testing 
the validity of each model. It follows that some of them are 
disfavored. Specifically, the QHD model, with the two different 
SU(6)-breaking mechanisms considered here~\citep{Close:2003wz}, 
systematically overestimates the {\tt NNPDF} result.
The statistical model~\citep{Bourrely:2001du} fails in the description 
of the ratio of polarized to unpolarized PDFs: indeed, it assumes zero 
gluon polarization at the initial input scale $Q^2=4$ GeV$^2$, 
while recent jet production data in polarized {\it pp}
collisions at RHIC~\citep{Adamczyk:2014ozi} have definitely pointed towards 
a positive gluon 
polarization~\citep{Nocera:2014gqa,deFlorian:2014yva}\footnote{The 
original analysis in Ref.~\citep{Bourrely:2001du} has been recently
revised~\citep{Bourrely:2014uha}, allowing for a nonzero gluon polarization 
at the initial input scale. A large gluon polarization, comparable
with that of Refs.~\citep{Nocera:2014gqa,deFlorian:2014yva}, is found in 
Ref.~\citep{Bourrely:2014uha}.}.
The RCQM~\citep{Isgur:1998yb} slightly overestimates the 
{\tt NNPDF} result for the neutron photoabsorption asymmetry $A_1^n$ 
in the $x$ region covered by experimental data, with which the {\tt NNPDF}
result is in good agreement. A substantial discrepancy is
also seen between the LSS(BBS)~\citep{Leader:1997kw} and the {\tt NNPDF}
predictions, the former always being larger than the latter.
A reasonable agreement is finally found between {\tt NNPDF} and both the 
NJL model~\citep{Cloet:2005pp} and the parameterization 
by Avakian {\it et al.}~\citep{Avakian:2007xa}, which explicitly included
subleading terms of the form $\ln^2(1-x)$ in the PDF parameterization.

\item The comparison between predictions obtained from global QCD analyses,
namely {\tt NNPDF} and {\tt DSSV08}, is interesting in two respects.
Concerning the ratio of polarized to unpolarized total $u$  
and $d$ quark combinations, the two parton sets are in perfect 
agreement at $x\lesssim 0.3$, while they are slightly different at 
$x\gtrsim 0.3$. Interestingly, for $d$ quarks, the {\tt NNPDF} prediction  
turns up to positive values around $x=0.75$, while the {\tt DSSV08} 
prediction remains negative. 
Concerning the ratio of polarized to unpolarized gluon, the {\tt NNPDF}
prediction is larger than the {\tt DSSV08} prediction. This is due
to the different behavior of the polarized gluon in the two parton sets: 
indeed, this is definitely positive in {\tt NNPDFpol1.1}, while it has a node 
in {\tt DSSV08}. The reason for this difference is that 
jet production data in polarized {\it pp} collisions were included in 
{\tt NNPDFpol1.1}, but were not in the 
original {\tt DSSV08} analysis. Actually, the latter has been recently
updated~\citep{deFlorian:2014yva} with the inclusion of these data, and a 
gluon polarization comparable to that in {\tt NNPDFpol1.1} was found.

 \item The possibility to discriminate between models at very large-$x$ values
is limited by the wide uncertainties which affect the {\tt NNPDF} predictions.
Indeed, all model expectations at $x\to 1$ provided in Tab.~\ref{tab:models}  
are compatible, within uncertainties, with the {\tt NNPDF} result at $x=0.9$
and $x=0.8$. At a more moderate value of $x$, $x=0.7$, uncertainties are 
well under control. This suggest that the behavior of PDFs remains largely 
uncertain at $x\gtrsim 0.7$, where no experimental data are available.
Furthermore, as the endpoint $x=1$ is approached, the accuracy of NLO
perturbative evolution is affected by powers of $\ln(1-x)$ which appear 
in the perturbative coefficients. Also nonperturbative effects,
like instantons or axial ghosts, may become relevant (see {\it e.g.} 
Sec.~9 of Ref.~\citep{Anselmino:1994gn}). 

\item The effective exponents $\beta_q$, defined by Eq.~(\ref{eq:betadef}), 
estimate the powerlike
behavior of PDFs at sufficiently large $x$ values, where 
the latter can be approximated as $q\sim (1-x)^{\beta_q}$; $q$
denotes either unpolarized or polarized distributions, respectively
$q=u,\bar{u},d,\bar{d},\bar{s},g$ or 
$q=\Delta u,\Delta\bar{u},\Delta d,\Delta\bar{d},\Delta\bar{s},\Delta g$. 
Results in Tab.~\ref{tab:prepexp-s} and in Fig.~\ref{fig:beta} 
suggest that the behavior of effective exponents 
for quark and antiquarks distributions
is consistent, within uncertainties, 
with the expectation based on QCD counting 
rules~\citep{Farrar:1975yb,Brodsky:1980ex}.
Indeed, these predict that, for a nucleon with helicity $+1/2$, 
\begin{eqnarray}
q        &\xrightarrow{x\sim 1}& (1-x)^{2n_s-1} + (1-x)^{2n_s+1}
\, \mbox{,}\\
\Delta q &\xrightarrow{x\sim 1}& (1-x)^{2n_s-1} - (1-x)^{2n_s+1}
\, \mbox{,}
\label{eq:countrules}
\end{eqnarray}
with $n_s$ the number of spectator quarks. Assuming $n_s=2$, it follows 
that the leading behavior of both unpolarized and polarized PDFs is 
$q\sim\Delta q \sim (1-x)^3$ as $x\to 1$. 
However, this behavior cannot hold at all $Q^2$, since
evolution causes the power of $(1-x)$ to grow like $\ln^2Q^2$ as $Q^2$
increases: this may explain the deviation from $\beta_q=3$
observed in Tab.~\ref{tab:prepexp-s} and Fig.~\ref{fig:beta}. 

\item At sufficiently large-$x$ values, the effective exponents for 
both unpolarized and polarized PDFs tend to coincide, and this means
that the LO positivity bound $|\Delta q|\leq q$, 
$q=u,\bar{u},d,\bar{d},\bar{s},g$ is saturated.
Such a behavior occurs at very large values of $x$, typically
$x\sim 0.85$.

\item A new determination of unpolarized PDFs based on the {\tt NNPDF}
methodology, {\tt NNPDF3.0}~\citep{Ball:2014uwa}, has been released recently. 
As in all {\tt NNPDF} determinations, the neural network parameterization of 
each PDF $i$ is supplemented by a preprocessing term of the form 
$x^{-a_i}(1-x)^{b_1}$; $a_i$ and $b_i$ are chosen, at random for each PDF replica,
in a given interval of values. In {\tt NNPDF2.3}, this interval was determined
based on a stability analysis of the results, while in {\tt NNPDF3.0} and 
{\tt NNPDFpol1.1} this is determined in an automatic and self-consistent way
(for details see Ref.~\citep{Ball:2014uwa}). In the future, it would  
be interesting to study whether the results in Tab.~\ref{tab:prepexp-s} and 
Fig.~\ref{fig:beta}, including the mutual spreads of the unpolarized and 
polarized large-$x$ effective exponents, will change upon the methodological 
improvement introduced in {\tt NNPDF3.0}.
 
\end{itemize}

The investigation of the large-$x$ behavior of polarized PDFs is one of the
goals pursued by ongoing and future JLAB experimental 
programs~\citep{Dudek:2012vr}. Several data sets on neutron and proton 
asymmetries in inclusive DIS, $A_1^n$ and $A_1^p$, have become 
available recently, specifically from JLAB E99-117~\citep{Zheng:2004ce}, 
JLAB E93-009~\citep{Dharmawardane:2006zd} and 
JLAB E06-014~\citep{Parno:2014xzb} experiments. 
These are shown in Fig.~\ref{fig:asy}, together with 
E143~\citep{Abe:1998wq}, E154~\citep{Abe:1997cx},
E155~\citep{Anthony:2000fn} and HERMES~\citep{Airapetian:2006vy} data sets,
but, at variance with the latter, they were not included in the 
{\tt NNPDFpol1.1} analysis. 
A large amount of data on the ratio of 
polarized to unpolarized proton and deuteron structure functions, 
$g_1^{p,d}/F_1^{p,d}$, has been also measured by CLAS~\citep{Prok:2014ltt}
very recently.

Qualitatively, JLAB data appear in agreement
with the {\tt NNPDF} prediction of the neutron and proton 
photoabsorption asymmetries, $A_1^n$ and $A_1^p$, in Fig.~\ref{fig:asy}.
Nevertheless, it is apparent that they are much more accurate than 
previous measurements already included in {\tt NNPDFpol1.1}.
In order to quantitatively assess the impact of JLAB data on 
PDFs, one should then include them in a global 
determination, with a careful treatment of dynamic higher-twist 
contributions to the Wilson expansion. 
Indeed, JLAB data are taken in a kinematic region (large $x$, small $Q^2$)
where the inclusion of these effects were shown to be essential
for describing them correctly~\citep{Jimenez-Delgado:2013boa}.

In a previous {\tt NNPDF} analysis~\citep{Ball:2013lla}, 
a conservative cut on the squared invariant 
mass of the hadronic final state $W^2=m^2+Q^2(1-x)/x$, 
with $m$ the nucleon target mass, 
was imposed in order to remove experimental data which may be
affected by sizable higher-twist corrections.    
The cut was set to $W^2\geq W^2_{\rm cut}=6.25$ GeV$^2$, because 
above this threshold higher-twist contributions become 
compatible with zero, when added to the observables with a coefficient
fitted to the data~\citep{Simolo:2006iw}. Following this choice, almost all
JLAB data are excluded, except few points in the 
JLAB E06-014~\citep{Parno:2014xzb} and CLAS~\citep{Prok:2014ltt} sets.
For this reason, the inclusion of JLAB data in a determination
of polarized PDFs based on the {\tt NNPDF} methodology is left for future
work. Of course, this should include a careful
treatment of dynamic higher-twist contributions along the lines of 
what has already been presented in the unpolarized case~\citep{Ball:2013gsa}.
}

\paragraph{Conclusions}
{
I have studied the behavior of polarized parton distributions in the regions of 
small and large momentum fractions, based on previous mutually consistent 
{\tt NNPDF} determinations of polarized~\citep{Nocera:2014gqa} 
and unpolarized~\citep{Ball:2012cx} PDFs. 
Among all PDF sets, these are the best suited in order for such a study 
to be effective: indeed, they include all the relevant experimental information
which is presently available, and they are determined with a methodology 
devised to provide a minimally biased result. 

I have investigated the potential of {\tt NNPDF} parton sets in discriminating 
the reliability of several theoretical models of polarized nucleon structure, by
comparing expectations for relevant observables based on them 
with the corresponding predictions obtained with {\tt NNPDF}. 
Only a limited number of models are clearly disfavored, 
while the possibility to discriminate between
the others is seriously limited by the large uncertainties which affect
the {\tt NNPDF} predictions at both small- and large-$x$ values.
  
The experimental information which will be provided either by a future 
high-energy polarized EIC~\citep{Accardi:2012qut} or 
by $12$ GeV JLAB upgrade~\citep{Dudek:2012vr}
will significantly reduce PDF uncertainties in global determinations,
at small and large $x$ respectively, 
and finally improve our knowledge of the nucleon spin structure in 
these regions. 
}

\section*{Acknowledgements}
I would like to thank S.~Forte, G.~Ridolfi and J.~Rojo for their 
useful comments to the manuscript. 
I also acknowledge the kind hospitality of the theory group at 
the Nationaal instituut voor subatomaire fysica (Nikhef), 
where this work was partly performed.
This research was supported by an Italian PRIN2010 grant and
by a European Investment Bank EIBURS grant.

\section*{References}
\bibliography{paper.bib}

\begin{thebibliography}{10}

\bibitem{Bass:2004xa}
S.D. Bass,
\newblock Rev.Mod.Phys. 77 (2005) 1257, hep-ph/0411005.

\bibitem{Kuhn:2008sy}
S. Kuhn, J.P. Chen and E. Leader,
\newblock Prog.Part.Nucl.Phys. 63 (2009) 1, 0812.3535.

\bibitem{Bjorken:1966jh}
J. Bjorken,
\newblock Phys.Rev. 148 (1966) 1467.

\bibitem{Ellis:1973kp}
J.R. Ellis and R.L. Jaffe,
\newblock Phys.Rev. D9 (1974) 1444.

\bibitem{Burkhardt:1970ti}
H. Burkhardt and W. Cottingham,
\newblock Annals Phys. 56 (1970) 453.

\bibitem{Efremov:1996hd}
A. Efremov, O. Teryaev and E. Leader,
\newblock Phys.Rev. D55 (1997) 4307, hep-ph/9607217.

\bibitem{Cabibbo:2003cu}
N. Cabibbo, E.C. Swallow and R. Winston,
\newblock Ann.Rev.Nucl.Part.Sci. 53 (2003) 39, hep-ph/0307298.

\bibitem{Hirai:2008aj}
M. Hirai and S. Kumano,
\newblock Nucl.Phys. B813 (2009) 106, 0808.0413.

\bibitem{deFlorian:2009vb}
D. de~Florian et~al.,
\newblock Phys.Rev. D80 (2009) 034030, 0904.3821.

\bibitem{Leader:2010rb}
E. Leader, A.V. Sidorov and D.B. Stamenov,
\newblock Phys.Rev. D82 (2010) 114018, 1010.0574.

\bibitem{Blumlein:2010rn}
J. Blumlein and H. Bottcher,
\newblock Nucl.Phys. B841 (2010) 205, 1005.3113.

\bibitem{Ball:2013lla}
The NNPDF Collaboration, R.D. Ball et~al.,
\newblock Nucl.Phys. B874 (2013) 36, 1303.7236.

\bibitem{Arbabifar:2013tma}
F. Arbabifar, A.N. Khorramian and M. Soleymaninia,
\newblock Phys.Rev. D89 (2014) 034006, 1311.1830.

\bibitem{Jimenez-Delgado:2013boa}
P. Jimenez-Delgado, A. Accardi and W. Melnitchouk,
\newblock Phys.Rev. D89 (2014) 034025, 1310.3734.

\bibitem{deFlorian:2014yva}
D. de~Florian et~al.,
\newblock Phys.Rev.Lett. 113 (2014) 012001, 1404.4293.

\bibitem{Nocera:2014gqa}
NNPDF Collaboration, E.R. Nocera et~al.,
\newblock Nucl.Phys. B887 (2014) 276, 1406.5539.

\bibitem{Nocera:2014vla}
E.R. Nocera,
\newblock (2014), 1403.0440.

\bibitem{Accardi:2012qut}
A. Accardi et~al.,
\newblock (2012), 1212.1701.

\bibitem{Dudek:2012vr}
J. Dudek et~al.,
\newblock Eur.Phys.J. A48 (2012) 187, 1208.1244.

\bibitem{Ball:2012cx}
The NNPDF Collaboration, R.D. Ball et~al.,
\newblock Nucl.Phys. B867 (2013) 244, 1207.1303.

\bibitem{Ball:2012wy}
R.D. Ball et~al.,
\newblock JHEP 1304 (2013) 125, 1211.5142.

\bibitem{Heimann:1973hq}
R. Heimann,
\newblock Nucl.Phys. B64 (1973) 429.

\bibitem{Bass:2006dq}
S.D. Bass,
\newblock Mod.Phys.Lett. A22 (2007) 1005, hep-ph/0606067.

\bibitem{Bass:1994xb}
S. Bass and P. Landshoff,
\newblock Phys.Lett. B336 (1994) 537, hep-ph/9406350.

\bibitem{Close:1994he}
F. Close and R. Roberts,
\newblock Phys.Lett. B336 (1994) 257, hep-ph/9407204.

\bibitem{Ball:1995ye}
R.D. Ball, S. Forte and G. Ridolfi,
\newblock Nucl.Phys. B444 (1995) 287, hep-ph/9502340,
\newblock Erratum-ibid. B449 (1995) 680.

\bibitem{Altarelli:1998nb}
G. Altarelli et~al.,
\newblock Acta Phys.Polon. B29 (1998) 1145, hep-ph/9803237.

\bibitem{Gehrmann:1995ut}
T. Gehrmann and W.J. Stirling,
\newblock Phys.Lett. B365 (1996) 347, hep-ph/9507332.

\bibitem{Ahmed:1975tj}
M. Ahmed and G.G. Ross,
\newblock Phys.Lett. B56 (1975) 385.

\bibitem{Bartels:1995iu}
J. Bartels, B. Ermolaev and M. Ryskin,
\newblock Z.Phys. C70 (1996) 273, hep-ph/9507271.

\bibitem{Bartels:1996wc}
J. Bartels, B. Ermolaev and M. Ryskin,
\newblock Z.Phys. C72 (1996) 627, hep-ph/9603204.

\bibitem{Ermolaev:2003zx}
B. Ermolaev, M. Greco and S. Troyan,
\newblock Phys.Lett. B579 (2004) 321, hep-ph/0307128.

\bibitem{Moch:2014sna}
S. Moch, J. Vermaseren and A. Vogt,
\newblock (2014), 1409.5131.

\bibitem{Brodsky:1989db}
S.J. Brodsky and I. Schmidt,
\newblock Phys.Lett. B234 (1990) 144.

\bibitem{Brodsky:1994kg}
S.J. Brodsky, M. Burkardt and I. Schmidt,
\newblock Nucl.Phys. B441 (1995) 197, hep-ph/9401328.

\bibitem{Ball:2013tyh}
The NNPDF Collaboration, R.D. Ball et~al.,
\newblock Phys.Lett. B728 (2014) 524, 1310.0461.

\bibitem{Aschenauer:2012ve}
E.C. Aschenauer, R. Sassot and M. Stratmann,
\newblock Phys.Rev. D86 (2012) 054020, 1206.6014.

\bibitem{Adeva:1998vv}
Spin Muon Collaboration, B. Adeva et~al.,
\newblock Phys.Rev. D58 (1998) 112001.

\bibitem{Abe:1998wq}
E143 collaboration, K. Abe et~al.,
\newblock Phys.Rev. D58 (1998) 112003, hep-ph/9802357.

\bibitem{Alekseev:2010hc}
COMPASS Collaboration, M. Alekseev et~al.,
\newblock Phys.Lett. B690 (2010) 466, 1001.4654.

\bibitem{Airapetian:2006vy}
HERMES Collaboration, A. Airapetian et~al.,
\newblock Phys.Rev. D75 (2007) 012007, hep-ex/0609039.

\bibitem{Bourrely:2001du}
C. Bourrely, J. Soffer and F. Buccella,
\newblock Eur.Phys.J. C23 (2002) 487, hep-ph/0109160.

\bibitem{Leader:1997kw}
E. Leader, A.V. Sidorov and D.B. Stamenov,
\newblock Int.J.Mod.Phys. A13 (1998) 5573, hep-ph/9708335.

\bibitem{Blumlein:1996hb}
J. Blumlein and A. Vogt,
\newblock Phys.Lett. B386 (1996) 350, hep-ph/9606254.

\bibitem{Close:1974ux}
F. Close,
\newblock Nucl.Phys. B80 (1974) 269.

\bibitem{Close:1973xw}
F. Close,
\newblock Phys.Lett. B43 (1973) 422.

\bibitem{Isgur:1998yb}
N. Isgur,
\newblock Phys.Rev. D59 (1999) 034013, hep-ph/9809255.

\bibitem{Close:2003wz}
F. Close and W. Melnitchouk,
\newblock Phys.Rev. C68 (2003) 035210, hep-ph/0302013.

\bibitem{Cloet:2005pp}
I. Cloet, W. Bentz and A.W. Thomas,
\newblock Phys.Lett. B621 (2005) 246, hep-ph/0504229.

\bibitem{Roberts:2013mja}
C.D. Roberts, R.J. Holt and S.M. Schmidt,
\newblock Phys.Lett. B727 (2013) 249, 1308.1236.

\bibitem{Wakamatsu:2014asa}
M. Wakamatsu,
\newblock (2014), 1405.7095.

\bibitem{Kochelev:1997ux}
N. Kochelev,
\newblock Phys.Rev. D57 (1998) 5539, hep-ph/9711226.

\bibitem{Boros:1999tb}
C. Boros and A.W. Thomas,
\newblock Phys.Rev. D60 (1999) 074017, hep-ph/9902372.

\bibitem{Farrar:1975yb}
G.R. Farrar and D.R. Jackson,
\newblock Phys.Rev.Lett. 35 (1975) 1416.

\bibitem{Avakian:2007xa}
H. Avakian et~al.,
\newblock Phys.Rev.Lett. 99 (2007) 082001, 0705.1553.

\bibitem{Martin:2002aw}
A. Martin et~al.,
\newblock Eur.Phys.J. C28 (2003) 455, hep-ph/0211080.

\bibitem{Wandzura:1977qf}
S. Wandzura and F. Wilczek,
\newblock Phys.Lett. B72 (1977) 195.

\bibitem{Adamczyk:2014ozi}
STAR Collaboration, L. Adamczyk et~al.,
\newblock (2014), 1405.5134.

\bibitem{Bourrely:2014uha}
C. Bourrely and J. Soffer,
\newblock (2014), 1408.7057.

\bibitem{Anselmino:1994gn}
M. Anselmino, A. Efremov and E. Leader,
\newblock Phys.Rept. 261 (1995) 1, hep-ph/9501369,
\newblock Erratum-ibid. 281 (1997) 399-400.

\bibitem{Brodsky:1980ex}
S.J. Brodsky and G.P. Lepage,
\newblock Phys.Scripta 23 (1981) 945.

\bibitem{Ball:2014uwa}
The NNPDF Collaboration, R.D. Ball et~al.,
\newblock (2014), 1410.8849.

\bibitem{Zheng:2004ce}
JLab Hall A Collaboration, X. Zheng et~al.,
\newblock Phys.Rev. C70 (2004) 065207, nucl-ex/0405006.

\bibitem{Dharmawardane:2006zd}
CLAS Collaboration, K. Dharmawardane et~al.,
\newblock Phys.Lett. B641 (2006) 11, nucl-ex/0605028.

\bibitem{Parno:2014xzb}
Jefferson Lab Hall A Collaboration, D. Parno et~al.,
\newblock (2014), 1406.1207.

\bibitem{Abe:1997cx}
E154 Collaboration, K. Abe et~al.,
\newblock Phys.Rev.Lett. 79 (1997) 26, hep-ex/9705012.

\bibitem{Anthony:2000fn}
E155 Collaboration, P. Anthony et~al.,
\newblock Phys.Lett. B493 (2000) 19, hep-ph/0007248.

\bibitem{Prok:2014ltt}
CLAS Collaboration, Y. Prok et~al.,
\newblock Phys.Rev. C90 (2014) 025212, 1404.6231.

\bibitem{Simolo:2006iw}
C. Simolo,
\newblock (2006), 0807.1501.

\bibitem{Ball:2013gsa}
The NNPDF Collaboration, R.D. Ball et~al.,
\newblock Phys.Lett. B723 (2013) 330, 1303.1189.

\end{thebibliography}

\end{document}